\documentclass[
twocolumn,
aps,
prb,
preprintnumbers,
amsmath,
amssymb,
superscriptaddress,
showpacs,
10pt]{revtex4-1}
\usepackage{CJK}
\usepackage{color}
\usepackage{graphicx}
\usepackage{subfig}
\usepackage{dcolumn}
\usepackage{bm}
\usepackage{multirow}
\begin{document}
\begin{CJK*}{GBK}{song} 
\preprint{}
\title{A first-principles study on the phonon transport in layered BiCuOSe}
\author{Hezhu Shao}
\email{hzshao@nimte.ac.cn}
\affiliation{Ningbo Institute of Materials Technology and Engineering, Chinese Academy of Sciences, Ningbo 315201, China}
\author{Xiaojian Tan}
\affiliation{Ningbo Institute of Materials Technology and Engineering, Chinese Academy of Sciences, Ningbo 315201, China}
\author{Guo-Qiang Liu}
\affiliation{Ningbo Institute of Materials Technology and Engineering, Chinese Academy of Sciences, Ningbo 315201, China}
\author{Jun Jiang}
\email{jjun@nimte.ac.cn}
\affiliation{Ningbo Institute of Materials Technology and Engineering, Chinese Academy of Sciences, Ningbo 315201, China}
\author{Haochuan Jiang}
\email{jianghaochuan@nimte.ac.cn}
\affiliation{Ningbo Institute of Materials Technology and Engineering, Chinese Academy of Sciences, Ningbo 315201, China}
\date{\today}
\begin{abstract}
First-principles calculations are employed to investigate the phonon transport of BiCuOSe. Our calculations reproduce the lattice thermal conductivity of BiCuOSe.
The calculated gr\"{u}neisen parameter is 2.4$\sim$2.6 at room temperature, a fairly large value indicating a strong anharmonicity in BiCuOSe, which leads to its ultralow lattice thermal conductivity.
The contribution to total thermal conductivity from high-frequency optical phonons, which are mostly contributed by the vibrations of O atoms, is larger than 1/3, remarkably different from the usual picture with very little contribution from high-frequency optical phonons.
Our calculations show that both the high group velocities and low scattering processes involved make the high-frequency optical modes contribute
considerably to the total lattice thermal conductivity.
In addition, we show that the sound velocity and bulk modulus along $a$ and $c$ axes exhibit strong anisotropy, which results in the anisotropic thermal conductivity in BiCuOSe.
\end{abstract}

\maketitle
\end{CJK*}

\section{Introduction}
Thermoelectric (TE) materials can realize direct energy conversion between heat and electricity, and have many potential applications in power generation and heat pumping~\cite{Rowe1995}.
TE performance is determined by a dimensionless figure of merit, $ZT=S^2\sigma T/\kappa$, where $S$, $\sigma$, $\kappa$, and $T$ are the Seebeck coefficient, electrical conductivity, thermal conductivity, and absolute temperature, respectively.
High TE performance can be achieved in materials with large Seebeck coefficient, high electrical conductivity, and low thermal conductivity.

Recently, layered BiCuOSe has attracted considerable interests in TE applications, because it exhibits very low intrinsic thermal conductivity~\cite{Liu2011,Barreteau2012,Li2012,Li2012a,Lan2013,Li2013,Sui2013,Lee2013,Pei2013,Barreteau2014,Li2014,Tan2014,Li2015,Liu2015,Liu2015a,Ren2015}.
In the past few years, significant effort has been made to improve the $ZT$ value of BiCuOSe. With doped various metal elements such as divalent Mg, Ca, Sr, Ba, Pb, and Zn~\cite{Barreteau2012,Li2012a,Li2013,Pei2013,Lan2013,Ren2015}, monovalent Na, K, and Ag~\cite{Lee2013,Li2014,Tan2014}, and trivalent La~\cite{Liu2015}, and also with Cu-deficient self-doping~\cite{Liu2011}, the electrical transport properties of BiCuOSe could be improved remarkably, due to the increase in the concentration of carries. Additionally, the anisotropic property of layered BiCuOSe can help to enhance the $ZT$ value further.
By hot-forging process, Sui {\it el al} obtained textured Ba-doped BiCuSeO with a high $ZT$ of 1.4 at 923 K~\cite{Sui2013}.
On the other hand, it was found that point defects arisen from doping~\cite{Barreteau2012,Li2012a,Li2013,Pei2013,Lan2013,Li2014,Tan2014,Ren2015} or vacancies in BiCuOSe~\cite{Liu2011,Li2015} could reduce its lattice thermal conductivity effectively.

Much research on the electronic structures and electronic carrier transport properties of BiCuOSe
has been done~\cite{Barreteau2012,Hiramatsu2008,Stampler2008,Zakutayev2011,Sallis2012,Zou2013}.
It is found that, similar to other good semiconductor TE materials, BiCuOSe has a narrow band gap of 0.75--0.8 eV~\cite{Hiramatsu2008,Stampler2008}.
In BiCuOSe, there is a mixture of heavy and light bands near the valence band maximum~\cite{Zou2013}, which is beneficial for its good TE performance,  for that the light bands lead to high carriers mobility, and thus the high electrical conductivity, whereas the heavy bands result into a steeper density of states near the Fermi level and thus to a higher Seebeck coefficient~\cite{Rowe1995}.

The remarkable advantage of BiCuOSe for TE applications is its extraordinarily low lattice thermal conductivity.
It is reported that the lattice thermal conductivity of non-textured BiCuOSe is as low as 0.52--0.9 Wm$^{-1}$K$^{-1}$ at 300 K~\cite{Liu2011,Li2012,Li2012a,Lan2013,Pei2013}, which is much lower than that in state of the art bulk TE materials of Bi$_2$Te$_3$ (1.6 Wm$^{-1}$K$^{-1}$ at 300 K~\cite{Goldsmid1961}) or PbTe (2.8 Wm$^{-1}$K$^{-1}$ at 300 K~\cite{Akhmedova2009}), and even lower than that in phonon-glass electron-crystal materials of clathrates or skutterudites~\cite{Sootsman2009}.
Recently, some studies on the origin of low thermal conductivity of BiCuOSe have been conducted. Pei {\it et al} have speculated that the weak chemical bonding and strong anharmonicity of bonding resulted in the low thermal conductivity~\cite{Pei2013}. Saha has conducted a comparison study on the lattice dynamics between BiCuOSe and LaCuOSe, and ascribed the low thermal conductivity of BiCuOSe to its constituent elements with high atomic mass~\cite{Saha2015}. However, the investigation on the phonon transport properties of BiCuOSe is still relatively rare. Recently, {\it ab initio} calculation of lattice thermal conductivity has been developed by combining the first-principles
calculations of interatomic force constants with solving the phonon Boltzmann transport equation iteratively~\cite{Esfarjani2008,LiWu2014}.
The first-principles calculations can accurately predict the phonon thermal conductivity without any assumption on phonon lifetimes, and
capture the transport properties of each phonon, and have been used in the investigations of phonon transport properties for many materials  ~\cite{LiWu2014,Chernatynskiy2015,Togo2015,Broido2007,Esfarjani2011,Li2012c,Tian2012a,Lindsay2013a,Lindsay2013,Ma2014,Li2014a,Lee2014,Cepellotti2015}.

In this work, we present a detailed investigation of phonon transport property for BiCuOSe to understand the physical mechanism of ultralow thermal conductivity and strong anisotropy in phonon transport. Our calculations reproduce the lattice thermal conductivity of BiCuOSe.
The gr\"{u}neisen parameter, which measures the anharmonicity of bonding, is 2.4$\sim$2.6 at room temperature, indicating a strong anharmonicity in BiCuOSe.
This leads to the low thermal conductivity. Moreover, we find that the high-frequency optical phonons of BiCuOSe contribute considerably to the total thermal conductivity, which is remarkably different from usual picture with little contribution from optical phonons. Additionally, our calculations show that the obvious anisotropy of phonon conduction originates from the anisotropic sound velocities in BiCuOSe. 

\section{Methodology}
The calculations are based on density functional theory method
in the generalized gradient approximation with the Perdew, Burke, and Ernzerhof functional (PBE)~\cite{Perdew1996}, as implemented in the Vienna {\it Ab initio} Simulation Package (VASP), which employs a plane-wave basis~\cite{Kresse1993,Kresse1996}.
The plane-wave energy cutoff is set to be 500.00 eV, and the electronic energy convergence is $10^{-8}$ eV. During relaxations, the force convergence for ions is $10^{-3}$ eV/{\AA}. The PBE functional overestimates the equilibrium volume, and then underestimates the bulk modulus and group velocity of phonons. The hybrid functionals could be more reliable to obtain the equilibrium volume~\cite{Bahers2015} and the bulk modulus. We then employ HSE06~\cite{Heyd2003,Heyd2006} to calculate the equilibrium lattice parameters and bulk modulus of BiCuOSe, then estimate the errors by the PBE calculations.

The phonon frequencies and velocities are calculated within the harmonic approximation.
Parlinski-Li-Kawazoe method, which is based on the supercell approach with finite displacement method as implemented in the Phonopy package~\cite{Parlinski1997,Togo2008}, is employed to obtain the phonon dispersion, phonon density of states, and group velocities for BiCuOSe.
To obtain convergent phonon property, in the calculations of harmonic interatomic force constants, a $3\times3\times2$ supercell of primitive cell containing 144 atoms is employed, and a $\Gamma$--centered $2\times2\times1$ Monkhorst-Pack k-point mesh is used to sample the irreducible Brillouin zone.

The phonon Boltzmann transport equation solved iteratively as implemented in ShengBTE~\cite{LiWu2014} is employed to study the phonon transport of BiCuOSe.
The phonon lifetime can be solved numerically with an iterative method. The zeroth-order solution corresponds to the relaxation time approximation.
The RTA can well describe the phonon conduction of some materials such as Si and Ge~\cite{Broido2007}, while it fails to predict the thermal conductivity of carbon-based materials such as graphene~\cite{Cepellotti2015}. We also inspect how accurate the prediction by RTA is for the thermal conductivity of BiCuOSe.

To get the three-phonon scattering matrix elements for calculating the phonon lifetimes, the third-order force constants should be calculated firstly.
In ShengBTE, the third-order force constants are calculated by a finite-difference supercell approach.
For the force calculations, a $3\times3\times2$ supercell of primitive cell containing 144 atoms is employed so that there is only negligible interaction between atoms in the center and at the boundary. To get convergent lattice thermal conductivity, we adopt the eighth nearest neighbors as the cutoff for third-order force interactions. And there is 716 supercells with displaced atoms needed to be calculated in VASP.
During the calculations of force constants, only the $\Gamma$ point is set for the k-point grid. And in the calculation of lattice thermal conductivity, a 15 $\times$ 15 $\times$ 9 q-point grid (180 inequivalent {\bf q} points) is employed. Apart from three-phonon processes, the contribution to scattering probabilities from isotopic disorder can also be calculated in ShengBTE.
After calculating the phonon lifetime, the lattice thermal conductivity can be obtained by
\begin{equation}\label{eq:2c16}
\kappa_{\alpha\beta}=\frac{1}{N k_B T^2 \Omega} \sum_{\lambda}f_0(\omega_\lambda) (f_0(\omega_\lambda)+1)(\hbar\omega_\lambda)^2v_\lambda^{\alpha}v_\lambda^{\beta}\tau_{\lambda}.
\end{equation}
where $\Omega$ is the volume of primitive cell, $\alpha$ and $\beta$ are the Cartesian components of $x$, $y$, or $z$, $k_B$ is Boltzmann constant, and $f_0(\omega_\lambda)$ is the Bose-Einstein distribution function.

\section{Results and Discussion}

\subsection{Crystal structure}

\begin{figure}[!htbp]
\centering
\includegraphics[width=4cm]{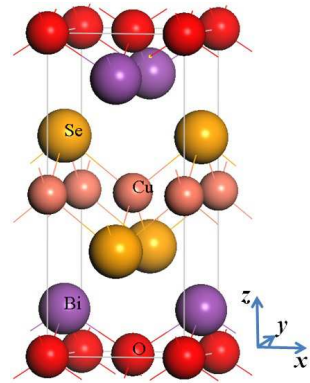}
 \caption{Crystal structure of BiCuSeO in tetragonal cell.}
 \label{figure:1}
 \end{figure}

 \begin{figure*}[!htbp]
\centering
\includegraphics[width=12cm]{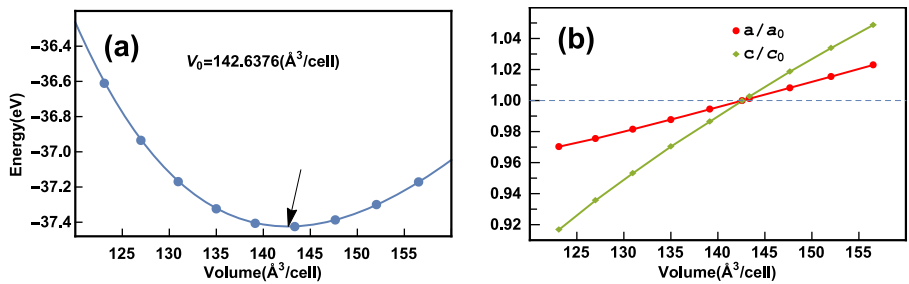}
 \caption{The structural optimization curves (a) and relative changing in lattice constants
 with volume (b) for BiCuOSe.}
 \label{figure:2}
 \end{figure*}

 \begin{table}[hbt]
\centering
\caption{Calculated equilibrium lattice constants (a and c in {\AA}), $z$ coordinates of Bi and Se atoms, bulk modulus (B in GPa),
and the pressure derivative of bulk modulus (B') along with the available experimental data at room temperature.}
\vspace{2pt}
\begin{tabular}{lllllll}
\hline
\hline
a  & c & $z$ (Bi) & $z$ (Se)& B & B' & Refs \\
\hline
3.956 & 9.116 & 0.139 & 0.672 & 72.5 & 5.16 & PBE \\
3.916 & 8.988 & 0.138 & 0.676 & 84.8 & 4.68 & HSE06 \\
3.928 & 8.933 & 0.140  & 0.674 &  &  & (Exp.) Ref~\cite{Barreteau2012} \\
3.935 & 8.937 &  & &  & & (Exp.) Ref~\cite{Lee2013} \\
\hline
\hline
\end{tabular}
\label{table:1}
\end{table}

BiCuOSe has tetragonal structure with space group of {\it P4/nmm}~\cite{Barreteau2012}, as shown in Fig.~\ref{figure:1}.
The unit cell of BiCuOSe contains eight atoms,
occupying four two-fold positions: O on 2a (0.75, 0.25, 0), Cu on 2b (0.75, 0.25, 0.5), Bi on 2c (0.25, 0.25, $z$ (Bi)), and
Se on 2c (0.25, 0.25, $z$ (Se)).
To determine the equilibrium lattice parameters, several volumes around the expected equilibrium volume are used to be relaxed for the ions and shapes.
And a set of volumes $V_i$ and energies $E_i$ are obtained. As shown in Fig.~\ref{figure:2}(a),
 we make a fit of these ($V_i$,  $E_i$)s to the Birch-Murnaghan 3rd-order equation of state~\cite{Birch1947}.
Table~\ref{table:1} presents the calculated lattice constants, atomic positions, bulk modulus, and pressure derivative of bulk modulus for BiCuOSe. We also plot the relative change of lattice constants with respect to volume in Fig.~\ref{figure:2}(b). It shows that in BiCuOSe, the $a$ axis is less sensitive to pressure or temperature than the $c$ axis.
The calculations are performed at zero temperature, and the experimental data are obtained at room temperature. Regardless of the thermal expansion, the HSE06 method gives a good prediction for the equilibrium volume within an error less than 0.4\%, while the PBE method overestimates the equilibrium volume of BiCuOSe with an error more than 3\%. However, now it is too expensive to perform the calculations of force constants by using the HSE06 method implemented in VASP, because the large amount of supercells with large number atoms are employed in the calculations. Then we use the PBE method to calculate the phonon property and thermal conductivity of BiCuOSe. Nevertheless, we could estimate roughly the errors of lattice thermal conductivity obtained by the PBE method. From Table~\ref{table:1}, the bulk modulus by PBE method is about 14\% less than that by the HSE06 method.
Because in solids, the velocity of elastic wave, $v \thicksim \sqrt{B}$. The lattice thermal conductivity $\kappa \thicksim v^2$. And then $\kappa \thicksim B$.
The PBE method may underestimate the lattice thermal conductivity by around 14\%.

\subsection{Phonon transport properties}

 \begin{table}[hbt]
\centering
\caption{The calculated dielectric constants and Born effective charges of BiCuOSe. Calculated data by PBEsol functional~\cite{Saha2015} for
comparison are given in parentheses.}
\vspace{2pt}
\begin{tabular}{ccc}
\hline
\hline
  & $xx$=$yy$ & $zz$ \\
\hline
$\varepsilon_{\infty}$ & 18.64 (18.01) & 14.01 (13.79)  \\
Z*(Bi) & 6.51 (6.46)& 6.05 (5.93)  \\
Z*(O) & $-$4.31 ($-$4.28)& $-$4.42 ($-$4.42) \\
Z*(Cu) & 1.52 (1.44)& 1.14 (1.06) \\
Z*(Se) & $-$3.73 ($-$3.67)& $-$2.76 ($-$2.70) \\
\hline
\hline
\end{tabular}
\label{table:2}
\end{table}

The calculated dielectric constants and Born effective charges are listed in Table~\ref{table:2}.
Our PBE calculations of dielectric constants and Born effective charges are slightly larger than those by PBEsol calculations of Saha~\cite{Saha2015}.
Because BiCuOSe has layered tetragonal structure with space group of {\it P4/nmm},
the Born effective charges and dielectric constants exhibit anisotropy with different in-plane and out-of-plane values. 
The dynamical matrix for calculation of phonon spectra of BiCuOSe can be constructed from the harmonic force constants by adding a non-analytical correction with dielectric constants and Born effective charges,
\begin{equation}\label{eq:2c2}
\tilde{D}_{\tau\alpha}^{\tau'\alpha'}({\bf q})=D_{\tau\alpha}^{\tau'\alpha'}({\bf q})+
\frac{4\pi}{\sqrt{M_{\tau} M_{\tau'}}\Omega} \frac{({\bf q}\cdot {\bf Z}_{\tau \alpha})({\bf q}\cdot {\bf Z}_{\tau' \alpha'}) }{{\bf q}^{T} {\bf \varepsilon_{\infty}} {\bf q}} e^{-\frac{{\bf q}^2}{\rho^2}},
\end{equation}
where ${\bf Z}$ and ${\bf \varepsilon}_{\infty}$ are the Born effective charge and dielectric constant, respectively.
And the dynamical matrix
\begin{equation}\label{eq:2c1}
D_{\tau \alpha}^{\tau' \alpha'}({\bf q})=\sum_{n}\frac{1}{\sqrt{M_{\tau}M_{\tau'}}}\Phi_{\tau\alpha}^{\tau'\alpha'}(n)\exp(i{\bf q}\cdot{\bf R}_n).
\end{equation}
where n denotes the $n$th primitive cell in supercell and ${\bf R}_n$ a translation vector for the $n$th primitive cell,  $\tau$ and $M_{\tau}$ refer to the atom and its mass in the primitive cell, $\alpha$ is the Cartesian components of $x$, $y$, or $z$, and $\Phi_{\tau\alpha}^{\tau'\alpha'}$ is the harmonic force constants.
The parameter $\rho$ (=0.25) is set so that the non-analytical term becomes negligible at the zone boundaries.

\begin{figure*}[!htbp]
\centering
\includegraphics[width=12cm]{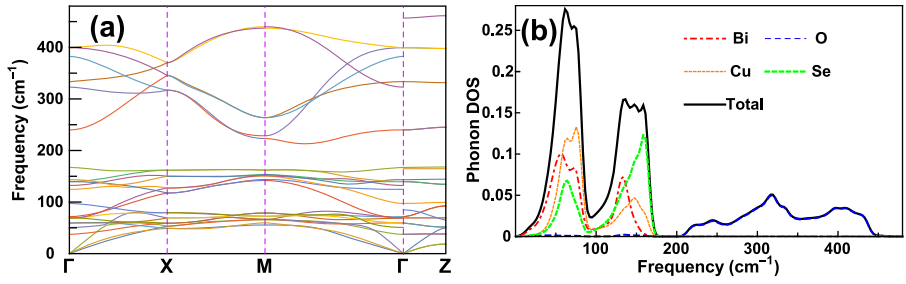}
 \caption{\label{figure:3} Phonon spectra (a) and phonon DOS (b) of BiCuOSe.}
 \end{figure*}

Fig.~\ref{figure:3} presents the phonon spectra along several high symmetry lines in Brillouin zone and phonon density of states.
The primitive cell of BiCuOSe contains eight atoms, and there are 24 phonon branches in the phonon spectra.
As shown in Fig.~\ref{figure:3}, the high-frequency optical modes are separated from the low-frequency modes by a gap of 46 cm$^{-1}$.
Because of the different directions of $\Gamma$-M and $\Gamma$-Z, the non-analytical term describing the LO-TO splitting effect bring different corrections to the phonon modes near $\Gamma$ point. Then the phonon spectrum seems to be not continuous at $\Gamma$ point.
From Fig.~\ref{figure:3}, the high-frequency optical modes above 213 cm$^{-1}$ are mainly from O vibrations, and these modes exhibit obvious dispersion, indicating that they have relatively large group velocities. While the acoustic and low-frequency optical modes are mainly from the vibrations of Bi, Cu, and Se atoms.

\begin{figure}[!htbp]
\centering
\includegraphics[width=6cm]{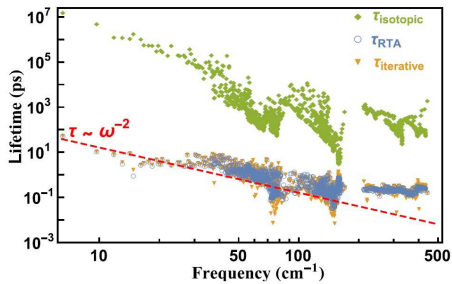}
 \caption{\label{figure:4} Frequency-dependent phonon lifetimes of  BiCuOSe at room temperature.}
 \end{figure}

The phonon lifetimes of BiCuOSe at room temperature are presented in Fig.~\ref{figure:4}. We calculate the three-phonon scattering rates and the contribution of isotopic disorder to scattering probabilities.
Fig.~\ref{figure:4} shows that compared to the anharmonic phonon-phonon scattering, the isotopic scattering (1/$\tau_{isotopic}$) is negligible to the thermal resistance.
There is a slight difference between the lifetimes by relaxation time approximation (RTA) and those of exact numerical solutions of the phonon Boltzmann equation, which means that the RTA describes the lattice thermal conductivity of BiCuOSe quite well. It implies that the umklapp scattering is dominant for the thermal resistance, and the redistribution of phonons resulting from the normal scattering processes has little influence on the phonon transport of BiCuOSe.
The lifetimes of most acoustic and optical phonons with frequencies lower than 60 cm$^{-1}$ are several ps', and for high-frequency optical modes, they lie in 0.1ps to 1ps in BiCuOSe. Such frequency-dependent lifetimes have the same magnitude with those of PbTe~\cite{Tian2012a}, while two orders of magnitude lower than those of Si~\cite{Esfarjani2011}.
Additionally, in the low-frequency range, the lifetimes show $\omega^{-2}$ dependence, which agrees with Klemens' prediction~\cite{Klemens1951}. While the lifetimes of high-frequency optical modes (above 213 cm$^{-1}$) behave one order of magnitude higher than those predicted by Klemens' relation.

\begin{figure}[!htbp]
\centering
\includegraphics[width=6cm]{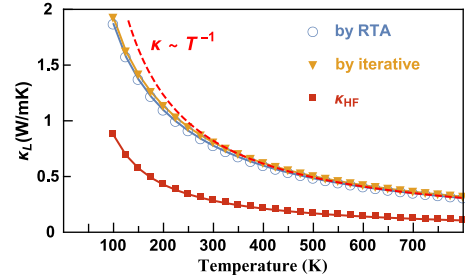}
 \caption{\label{figure:5} Calculated average lattice thermal conductivities of BiCuOSe with respect to temperature, $\kappa_{HF}$ denotes the contributions to total lattice thermal conductivity from the high-frequency (above 213 cm$^{-1}$) optical modes.}
 \end{figure}

Fig.~\ref{figure:5} shows the average temperature-dependant lattice thermal conductivity of BiCuOSe from 100 to 800 K, and as well as the contributions from high-frequency optical modes to the total lattice thermal conductivity. The average lattice thermal conductivity is calculated by $\kappa_L=(\kappa_{xx}+\kappa_{yy}+\kappa_{zz})/3$. As shown in Fig.~\ref{figure:5}, the thermal conductivities of BiCuOSe predicted by RTA are slightly lower than those of iterative solutions of the phonon Boltzmann equation.
The calculated $\kappa_L$ at 300 K is 0.8 Wm$^{-1}$K$^{-1}$, which is consistent with experimental results~\cite{Liu2011,Li2012,Li2012a,Lan2013,Pei2013}.
We note here that the PBE calculations may underestimate the thermal conductivity by around 14\%. On the other hand, we merely consider the intrinsic phonon-phonon scattering, and ignore other scattering source such as crystal interfaces, and may overestimate the thermal conductivity.
Although BiCuOSe exhibits ultralow intrinsic thermal conductivity,  the lattice thermal conductivities obey the $\sim1/T$ relation at high temperature (above Debye temperature $\sim$ 240 K).
In addition, we find that the contribution of high-frequency (above 213 cm$^{-1}$) optical phonons, which are mostly from vibrations of O atom, to overall lattice thermal conductivity is larger than 1/3.  This is remarkably different from the usual picture with very little contribution of thermal conductivity from such high-frequency phonons in most bulk materials~\cite{Beechem2010}.

\begin{figure}[!htbp]
\centering
\includegraphics[width=6cm]{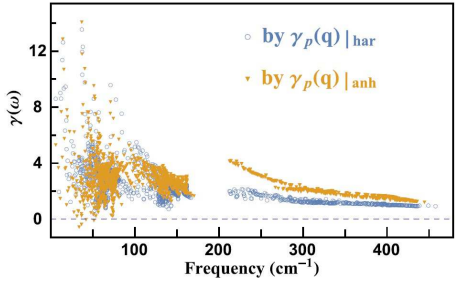}
 \caption{\label{figure:6} Frequency-dependent mode Gr\"{u}neisen parameters of BiCuOSe.}
 \end{figure}

To explore the mechanism of ultralow thermal conductivity, we continue to discuss the strength of phonon-phonon scattering processes.
At high temperatures, the phonon lifetime in crystal is determined by phonon-phonon scattering processes.
A useful measurement of the strength of phonon-phonon scattering is the mode Gr\"{u}neisen parameter $\gamma_p({\bf q})$, which can be calculated by
\begin{equation}\label{eq:3b1}
\begin{split}
&\gamma_p({\bf q}){\Big |}_{anh}=-\frac{1}{6\,\omega_{p}({\bf q})^2}\sum_{\eta' l'}\sum_{\eta'' l''}\sum_{\alpha \beta \gamma} \phi_{\eta 0,\eta' l',\eta'' l''}^{\alpha \beta \gamma}\\
&\times \frac{e_{\alpha \eta}^p({\bf q^*})e_{\beta \eta'}^p({\bf q})}{\sqrt{M_\eta M_{\eta'}}} e^{i {\bf q} \cdot {\bf R}_l'} {\bf r}_{\eta'' l'' \gamma},
\end{split}
\end{equation}
where $\phi_{\eta 0,\eta' l',\eta'' l''}^{\alpha\beta\gamma}$ are the third-order force constant, $e$ the phonon eigenvectors, and ${\bf r}_{\eta l}$ the position vector of the $\eta$th atom in $l$th primitive cell. On the other hand, one could obtain the mode Gr\"{u}neisen parameter by the definition of the phonon frequency shift with respect to the volume
\begin{equation}\label{eq:3b2}
\gamma_{p}({\bf q}){\Big |}_{har}=-\frac{V_0}{\omega_p({\bf q})}\frac{\partial \omega_p(\bf{q})}{\partial V},
\end{equation}
where $V_0$ is the equilibrium volume. Fig.~\ref{figure:6} presents the mode Gr\"{u}neisen parameters with respect to frequencies calculated by Eq.~(\ref{eq:3b1}) and (\ref{eq:3b2}) for BiCuOSe.
Eq.~(\ref{eq:3b1}) and  (\ref{eq:3b2}) give consistent results for the mode Gr\"{u}neisen parameters of acoustic and low-frequency optical phonons. While for high-frequency optical phonons, Eq.~(\ref{eq:3b1}) predicts some larger mode Gr\"{u}neisen parameters than those calculated by Eq.~(\ref{eq:3b2}). And the discrepancy between them becomes less and less with increased frequencies.

As shown in Fig.~\ref{figure:6}, throughout the Brillouin zone, the mode Gr\"{u}neisen parameters of BiCuOSe are mostly positive. Moreover the maxima of mode Gr\"{u}neisen parameters are very high near $\Gamma$ point and at low frequency range, up to 14. We further average $\gamma_p({\bf q})$ to obtain the average Gr\"{u}neisen parameter with temperatures
\begin{equation}\label{eq:3b3}
\gamma_{\rm{ave}}=\frac{1}{C_v} \sum_{p,{\bf q}}\gamma_{p}({\bf q})C_{v,p}({\bf q}),
\end{equation}
where $C_{v,n}(\mathbf{q})$ is the mode heat capacity and
\begin{equation}
C_v=\sum_{p,\mathbf{q}} k_B\left(\frac{\hbar\, \omega_p(\mathbf{q})}{k_B T}\right)^2 \frac{e^{\hbar\, \omega_p(\mathbf{q})/k_B T}}{(e^{\hbar\, \omega_p(\mathbf{q})/k_B T}-1)^2}.
 \label{eq:3b4}
\end{equation}
At room temperature, the obtained $\gamma_{\rm{ave}}$ is 2.4 and 2.6 by averaging the $\gamma_{p}({\bf q}){\Big |}_{har}$ and $\gamma_{p}({\bf q}){\Big |}_{anh}$ respectively. The calculated $\gamma_{\rm{ave}}$ of BiCuOSe is larger than that of PbTe, of which the $\gamma_{\rm{ave}}$ is 1.96$\sim$2.18, which was also obtained by first-principles calculations~\cite{Zhang2009}.  It is known that PbTe has very strong anharmonic phonon scattering~\cite{Bozin2010,Delaire2011}. Therefore, BiCuOSe behaves even more anharmonic than PbTe, and this leads to its ultralow intrinsic phonon conductivity.

\begin{figure}[!htbp]
\centering
\includegraphics[width=6cm]{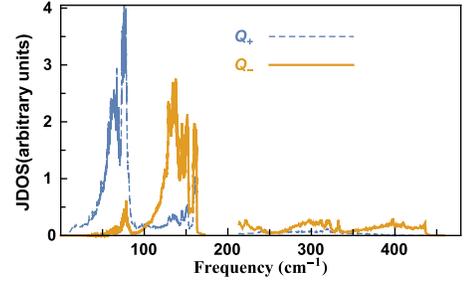}
 \caption{\label{figure:7} Frequency-dependent JDOS for scattering rate in BiCuOSe.}
 \end{figure}

In common materials, the high-frequency optical phonons exhibit small group velocities and suffer large scattering, thus make little contribution to the thermal conductivity. In contrast to usual picture, the high-frequency optical modes of BiCuOSe make remarkable contribution to the total thermal conductivity. To compare the frequency-dependant rates of scattered by other phonons in three-phonon processes, we define the joint density of state (JDOS) as
\begin{equation} \label{eq:3b5}
\begin{split}
&Q_{\pm}(\omega)=\sum_{p,p',p''}\delta(\omega-\omega_p({\bf q}))\\ &\int\int_{BZ} \delta(\omega_p({\bf q})\pm\omega_p'({\bf q'})-\omega_p''({\bf q}\pm{\bf q'}-{\bf G}))d{\bf q}d{\bf q'}.
\end{split}
\end{equation}
The JDOS is a representative of the phase space available for scattering events. $Q_+$ corresponds to the absorption processes, and $Q_-$ is for the emission processes. We present the JDOS of BiCuOSe in Fig.~\ref{figure:7}, which shows that both absorption and emission processes dominate for low-frequency phonons, while the high-frequency phonons involve much less scattering. Additionally, the high-frequency phonons are mainly from the vibrations of O atom. Due to the light atomic mass and strong interaction with Bi, the vibrations modes related to O atom exhibit strong dispersion, and their frequency range expands from 210 cm$^{-1}$ to 460 cm$^{-1}$. These high-frequency modes have relatively high group velocities.  Therefore, the high-frequency optical phonons make considerable contribution to the total thermal conductivity in BiCuOSe.

\subsection{Anisotropy of the thermal conductivities}

\begin{figure}[!htbp]
\centering
\includegraphics[width=6cm]{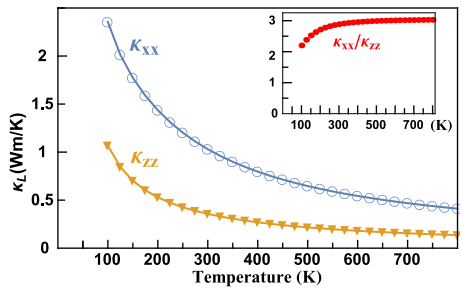}
 \caption{\label{figure:8} Temperature-dependent $\kappa_{xx}$ and $\kappa_{zz}$ of BiCuOSe. The inset of the figure shows the ratio of $\kappa_{xx}$ to $\kappa_{zz}$.}
 \end{figure}

Fig.~\ref{figure:8} presents the temperature-dependent lattice thermal conductivity along $a$ ($\kappa_{xx}$) and $c$ ($\kappa_{zz}$) axes of BiCuOSe, which shows strong anisotropy. The ratio of $\kappa_{xx}$ to $\kappa_{zz}$ is 2.3 $\sim$ 3. Recently, Saha has proposed that the different Gr\"{u}neisen parameters between in-plane and out-of-plane directions resulted in the anisotropy of thermal conductivity~\cite{Saha2015}.
However, in Slack's expression~\cite{Slack1979a}, which serves as estimation of lattice thermal conductivity, $\kappa_L=A\frac{\overline{M}\theta_D^{3}\delta}{\gamma^2n^{2/3}T}$, the square of Gr\"{u}neisen parameter $\gamma$ in the denominator is from the estimation of phonon relaxation lifetime, which is not direction-dependent.

\begin{figure}[!htbp]
\centering
\includegraphics[width=6cm]{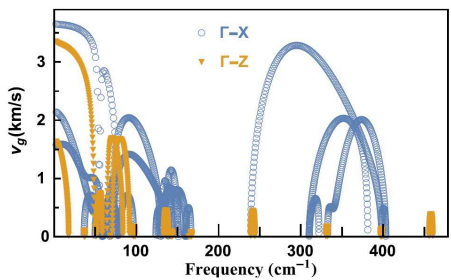}
 \caption{\label{figure:9} Group velocities along $\Gamma$-X and $\Gamma$-Z.}
 \end{figure}
To clarify the mechanism of anisotropy of the thermal conductivities, we firstly discuss the anisotropy of group velocity (GV) in BiCuOSe.
Fig.~\ref{figure:9} presents the group velocities along $\Gamma$-X and $\Gamma$-Z, which shows that the GVs along $\Gamma$-X are much larger than those along $\Gamma$-Z, especially for high-frequency optical modes.
The GV of longitudinal acoustic (LA) branch along $\Gamma$-X near zone center is about 3650 m/s, and the GV of transverse acoustic (TA) branches are 2138 and 1576 m/s. Whereas the GVs of LA and TA along $\Gamma$-Z near zone center are about 3354 and 1634 m/s. For high-frequency optical modes, the maximum GV along $\Gamma$-X reaches 3284 m/s, while that along $\Gamma$-Z is only 442 m/s.
Therefore, the GVs along in-plane and out-of-plane directions are very different.

\begin{table}[hbt]
\centering
\caption{Elastic constants of BiCuOSe. All values are in units of GPa.}
\vspace{2pt}
\begin{tabular}{llllll}
\hline
\hline
$c_{11}$  & $c_{33}$  & $c_{12}$  & $c_{13}$   & $c_{44}$  & $c_{66}$ \\
\hline
137.27&  94.54& 57.71 & 53.51 & 23.81 & 38.87   \\
 \hline
 \hline
\end{tabular}
\label{table:3}
\end{table}
Because $\kappa \thicksim B$, to give a quantity estimation for the ratio of in-plane to out-of-plane GV, we calculate the bulk modulus $B$ along $a$ and $c$ axes. The elastic constants are calculated and tabulated in Table~\ref{table:3}. The elastic constants satisfy all the Born's mechanical stability criteria, indicating that the system of BiCuOSe is in a mechanical stable state.
Using calculated elastic constants, the bulk and Young's modulus can be obtained by Voigt-Reuss-Hill approximations~\cite{Hill1952}.
The obtained bulk modulus from the elastic constants is 76.2 GPa, which is very close to that calculated from fitting the Birch-Murnaghan's 3rd-order equation of state.
The Young's modulus is a measure of the stiffness of an elastic material, and reflects the strength of bonding in the covalent crystals.
Our calculated Young's modulus of BiCuOSe is $E=79.6 $ GPa, which agrees well with the experimental results
of $E=76.5$ GPa by Pei {\it et al.}~\cite{Pei2013}. And the relatively small Young's modulus indicates a totally weak covalent bonding in BiCuOSe.
As is known, the elastic constants $c_{11}$ and $c_{33}$  reflect the stiffness-to-uniaxial strains along the
crystallographic $a$ and $c$ axes,  respectively,
while the elastic constants $c_{12}$, $c_{13}$, $c_{44}$, and $c_{66}$ are related to the elasticity in shape.
From Table~\ref{table:3}, the value of $c_{11}$ is much higher than $c_{33}$, which means that
BiCuOSe is stiffer for strains along the $a$ axis than along the $c$ axis.
Furthermore, the ratio of bulk modulus along $a$ axis to that along $c$ axis can be calculated by~\cite{Ravindran1998}
\begin{equation}
\frac{B_a}{B_c}=\frac{(c_{22}-c_{12})(c_{11}-c_{13})-(c_{11}-c_{12})(c_{23}-c_{12})}
{(c_{22}-c_{12})(c_{33}-c_{13})-(c_{12}-c_{23})(c_{13}-c_{23})}.
\end{equation}
The obtained $B_a/B_c$=2.3 for BiCuOSe, which shows strong anisotropy in elastic property. And this value is close to the ratio of $\kappa_{xx}$ to $\kappa_{zz}$.
Therefore, our calculations show that the strong anisotropy of lattice thermal conductivity in BiCuOSe should be ascribed to its anisotropic velocities.
The anisotropy of bulk modulus, which leads to anisotropic sound velocity in BiCuOSe, reflects the layered structure of BiCuOSe. And the relationship between such strong anisotropy and the layered structure and anisotropically chemical bonding in BiCuOSe remains to be uncovered.

\section{Conclusion}

In summary, we have employed first-principles calculations to investigate the phonon transport of BiCuOSe.
The obtained phonon lifetime of BiCuOSe is mostly in the range from 0.1 ps to several ps, which is comparable to those of PbTe.
Our calculations show that there is strong bonding anharmonicity in BiCuOSe, which leads to its ultralow lattice thermal conductivities.
Due to the high group velocities and low scattering processes involved, the high-frequency optical modes make a considerable contribution to the total lattice thermal conductivity. The anisotropy of thermal conductivities is ascribed to the anisotropy of wave velocities along $a$ and $c$ axes, which may originate from the layered structure and anisotropic chemical bonding in BiCuOSe. Our work facilitates deep understanding of the phonon transport properties in such layered materials.

\begin{acknowledgments}
H.S. acknowledges helpful discussions with Prof. Hao Zhang from Fudan University.
This work was supported by the National Natural Science Foundation of China (11404348, 11404350, and 11234012), Ningbo Municipal Natural Science Foundation (2014A610008), and Ningbo Science and Technology Innovation Team (2014B82004).
\end{acknowledgments}


\end{document}